\documentclass{article}

\usepackage{graphicx}          
\usepackage{amsmath,amssymb}
\usepackage{ascmac}

\title{\LARGE \bf 
Monte-Carlo Simulation of J1 League Postseason System from the 2015 Season
}

\author{Takeshi Izumi$^{1}$ and Eiji Konaka$^{1*}$
\thanks{$^{1}$Department of Information Engineering, 
Meijo University,
1-501 Shiogamaguchi, Tenpaku-ku, Nagoya, JAPAN.
        {*\tt\small konaka@meijo-u.ac.jp}}%
}

\begin{document}

\maketitle

\begin{abstract}

In this paper, the new two stage and postseason system of J1 League, the top division of professional football in Japan, is simulated.
Official regulation defines that the new postseason from the 2015 season consists of five teams selected by different principles --- the top three teams by total season points, and the winning teams of the 1st and 2nd half of the season.

This paper clarifies that there are overlaps within these five teams and the average number of teams reaching the postseason is about 3.5.
The probability that the postseason is held with 3, 4, and 5 teams are estimated about 0.62, 0.35, and 0.03, respectively.

From this result, this paper concludes that the new system of J1 League is basically a one-stage system, whereas the new system is officially defined as a two-stage system, due to inappropriate design of the postseason.
\end{abstract}

\section{Introduction}
J1 League, the top professional division of football in Japan, will adopt a new two-stage system with a postseason system from the 2015 season.
The postseason system consists of two different stages.
The (tentative) names are ``Super Stage" and ``J League Championship"\cite{JPlayoff2015}.

Since its opening in 1993 till 2004, the J1 League has had a system of two stages along with a chapmionship game.
From the 2005 season, the J1 League has adopted a simple double round-robin system like European major football leagues such as Premier League (England), Bundesliga (Germany), Serie A (Italy), Liga Espa\~nola (Spain), and so on.
Some European football leagues such as Jupiler Pro League (Belgium) and Scottish Premiership (Scotland) have a postseason system called the ``split system"\cite{JupilerProLeaguePlayoffSystem}\cite{SPFLRules}, but these are considered exceptional cases in UEFA\footnote{Union des Associations Europeennes de Football }.

The double round-robin principle is the most simple and fair system, however, J1 League has decided to change the system due to decreased sales and profit\cite{JPlayoff2015ChairmanInterview}.
\begin{quote}
Kazumi Ohigashi, the fourth Chairman of J League, said: 
``It is obvious that football in Japan has shown big progess and has been dramatically growing as a major sport in the last two decades.
Japanese players also have shown big progress.
Some Japanese play in top-level clubs in the world.
This is an amazing fact."

``However, money and good players are attracted to the five major European football leagues  (England, Spain, Italy, Germany, and France).
This is called the straw effect, and it refers to football leagues in other countries declining as playsers and money move to the major leagues; This s now occurring all over the world."

``J League income peaked at 12.8 billion yen in 2008, but decreased by 0.93 billion yen to 11.9 billion yen by 2012.
If this continues, income is expected to decline by another 1 billion yen by 2014."
(translated by the authors)
\end{quote}


Most professional sports league in the USA, such as NFL\footnote{National Football League}, MLB\footnote{Major League Baseball}, NBA\footnote{National Basketball Association}, and NHL\footnote{National Hockey League}, have adopted postseason  system.
These postseason systems are designed to get much attention, and these postseason games produce much profit.
NPB\footnote{Nippon Professional Baseball}, the top professional baseball league in Japan, has adopted a postseason system since 2007, which has seen some commercial success.
Following these successful examples, 
J1 League aims to increase its income by introducing the new postseason system. 

The most fundamental difference between the new postseason system of J1 League and that of others is that there is possibly an overlap in the teams which have the right to reach the postseason.
No substitution teams are selected even if there is any overlap, and therefore, 
the number of teams and games in the postseason may be different every year, although
the number of games is an important factor for income from box office sales and broadcast rights sales.

The objective of this paper is to simulate the new postseason system using a Monte-Carlo simulation to calculate the possibility of overlap.

\section{J1 League's new postseason system}
This section introduces the new postseason system.
The following explanation is based on the official press release in 2014/2/25\cite{JPlayoff2015}.

\begin{figure}[h]
\begin{center}
	\includegraphics[width=7.5cm ,clip]{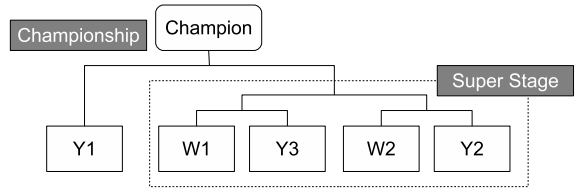}
	\caption{Postseason tournament of J1 League from the 2015 season}
	\label{fig:JPlayoff2015}
\end{center}
\end{figure}

\begin{figure}[h]
\begin{center}
	\includegraphics[width=9.0cm ,clip]{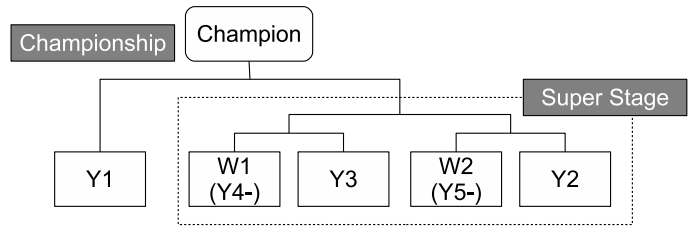}
	\caption{Overlap case \#1}
	\label{fig:JPlayoffCase1}
\end{center}
\end{figure}

\begin{figure}[h]
\begin{center}
	\includegraphics[width=9.0cm ,clip]{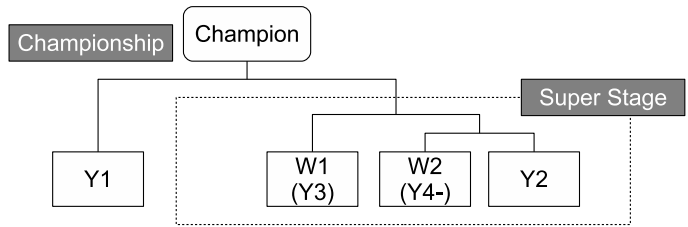}
	\caption{Overlap case \#2}
	\label{fig:JPlayoffCase2}
\end{center}
\end{figure}

\begin{figure}[h]
\begin{center}
	\includegraphics[width=9.0cm ,clip]{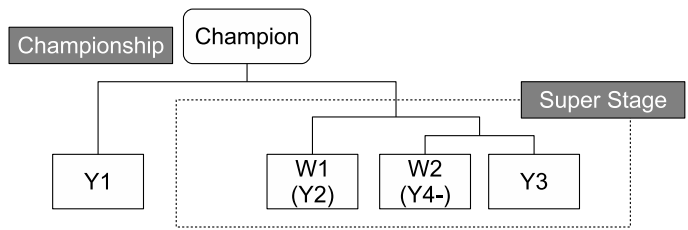}
	\caption{Overlap case \#3}
	\label{fig:JPlayoffCase3}
\end{center}
\end{figure}

\begin{figure}[h]
\begin{center}
	\includegraphics[width=9.0cm ,clip]{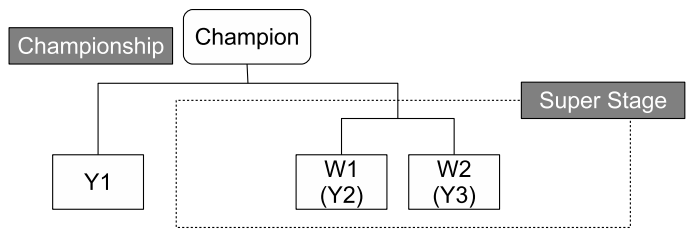}
	\caption{Overlap case \#4}
	\label{fig:JPlayoffCase4}
\end{center}
\end{figure}

\begin{figure}[h]
\begin{center}
	\includegraphics[width=9.0cm ,clip]{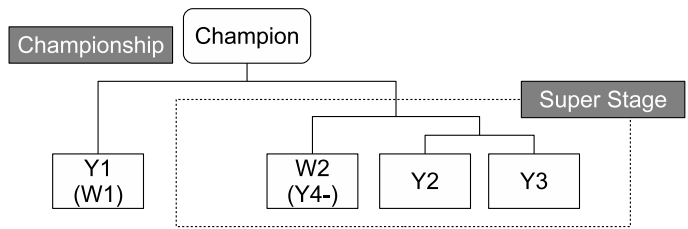}
	\caption{Overlap case \#5}
	\label{fig:JPlayoffCase5}
\end{center}
\end{figure}

\begin{figure}[h]
\begin{center}
	\includegraphics[width=9.0cm ,clip]{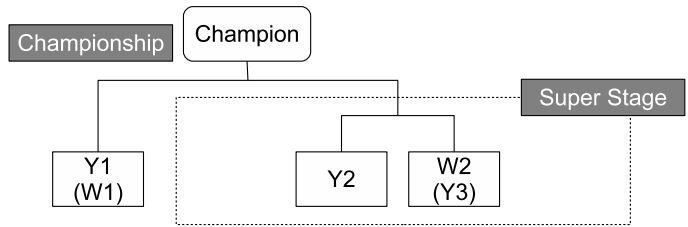}
	\caption{Overlap case \#6}
	\label{fig:JPlayoffCase6}
\end{center}
\end{figure}

\begin{figure}[h]
\begin{center}
	\includegraphics[width=9.0cm ,clip]{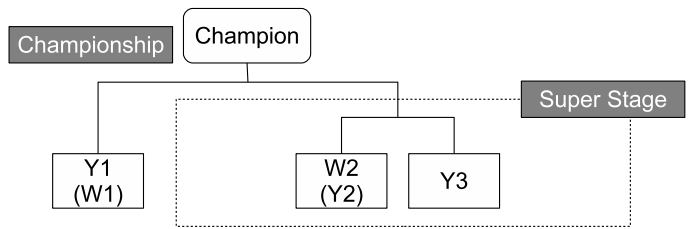}
	\caption{Overlap case \#7}
	\label{fig:JPlayoffCase7}
\end{center}
\end{figure}

\begin{figure}[h]
\begin{center}
	\includegraphics[width=9.0cm ,clip]{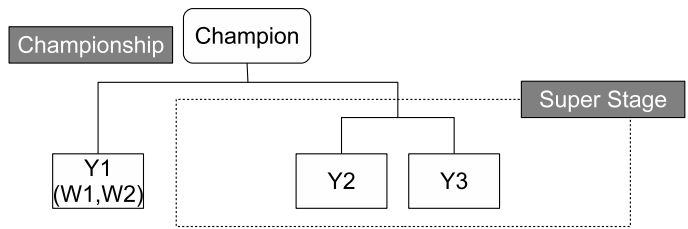}
	\caption{Overlap case \#8}
	\label{fig:JPlayoffCase8}
\end{center}
\end{figure}

\begin{itemize}
\item Regular season
	\begin{itemize}
	\item Ragular season consists of two stages by 18 clubs.
	Each stage is named as ``1st stage" and ``2nd stage."
	\item Each stage is single round-robin competition. Home and away games are placed on either stage. 
	As a result, each team has 17 matches in one stage with 8 or 9 home games.
	\item Standings for each stage are based on the sum in points of each game.
	The point system is the usual 3-1-0 (Win-Draw-Lose) system.
	\end{itemize}
\item Super Stage (tentative name)
	\begin{itemize}
	\item 
	The Super Stage is a knockout tournament of four teams.
	The following four teams reach the Super Stage.
		\begin{itemize}
		\item Winning teams of the 1st and 2nd stages.
		\item The teams ranked 2nd and 3rd in terms of  total points over the two stages.
		\end{itemize}
		The stage winning teams are denoted by W1 and W2 based on {\bf their total points over both stages}, not based on the stage number.
		Similarly, the latter teams are denoted by Y2 and Y3.
	\item Without any overlaps (defined below), match-ups in the first round are W1 and Y3, and W2 and Y2.
	The game is held in the hometown of the stage-winning team.
	\item The winners of the first round reach the second round, and the winner of the second round (a single-game contest) is the winner of the Super Stage. 
This team reaches the Championship.
	
	\item Overlap resolution
	\begin{itemize}
	\item If Y1 (defined below) overlaps with W1, Y1 reaches the Championship and does not play in the Super Stage.
	There is no subsutitution for W1.
	In this case, the match-up of the first stage is between Y2 and Y3. 
W2 is seeded and clears the first round.
	\item If either of W1 or W2 overlaps with Y2 or Y3, the team is seeded and clears the first round.
	\item If both W1 and W2 overlap Y2 and Y3, no first round is held and only the second-round match between W1(Y2) and W2(Y3) is held.
	\item If a stage-winning team is ranked from 16th to 18th in terms of total points over the two stages, the team does not reach the Super Stage\footnote{This condition is ignored in order to simplify the discussion. }.
	\end{itemize}
	\end{itemize}
	
\item J League Championship (tentative name)
	\begin{itemize}
	\item The team ranked 1st in terms of total points ovet the two stages, that is, over 34 matches, reaches  the J League Championship. This team is denoted as Y1.
	\item Y1 and the winner of the Super Stage play fot the Championship.
	The Championship consists of two games, one each in the hometowns of both teams.
	\end{itemize}
\end{itemize}
Figure \ref{fig:JPlayoff2015} shows the postseason tournament if five teams reach the postseason.
There are 8 possible cases of overlap (listed in Table \ref{tab:overlapCases}), and the number of teams reaching the postseason may vary from three to five.
Figures \ref{fig:JPlayoffCase1} to \ref{fig:JPlayoffCase8} show the tournament for each cases.
``Y4-" means that the stage-winning team is ranked 4th or lower in terms of the total points.

\begin{table}[h]
\begin{center}
\caption{Overlap cases}
\label{tab:overlapCases}
\begin{tabular}{clc}\\ \hline
case {\#}&overlap(s)& teams\\ \hline
1&null	&5\\ \hline
2&(Y3, W1) 	&4\\ \hline
3&(Y2, W1)	&4\\ \hline
4&(Y2, W1), (Y3, W2)	&3\\ \hline
5&(Y1, W1)	&4\\ \hline
6&(Y1, W1), (Y3, W2)	&3\\ \hline
7&(Y1, W1), (Y2, W2)	&3\\ \hline
8&(Y1, W1), (Y1, W2)	&3\\ \hline
\end{tabular}

\end{center}
\end{table}


\section{Simulation Results}
Intuitively, the stage-winning teams would rank high in terms of total points, and vice versa.
The objective of this paper is to simulate the new postseason decision system using Monte-Carlo simulation to calculate the possibility of an overlap.

\subsection{Notations}
The notations used in this paper are summarized below.

\begin{itemize}
\item $i$ and $j$ denote the  $i$-th and $j$-th teams, respectively. 
\item $\lambda_{ALL}$ denotes the average number of goals per game.
\item $\lambda_{i,GF}$	denotes the average number of goals per game by team $i$.
\item $\lambda_{i,GA}$ denotes the average number of goals per game against team $i$.
\item $\lambda_{i,GF,H}$ and $\lambda_{i,GF,A}$	denote the number average of goals per game by team $i$ in home and away games, respectively.
$\lambda_{i,GA,H}$ and $\lambda_{i,GA,A}$ are similarly defined.
\item $X_i$ is a random variable that shows the number of goals by team $i$ in one game.
\item $Po(\lambda)$ is a Poisson distribution with mean $\lambda$.
\end{itemize}

\subsection{Game model}
\begin{figure}[h]
\begin{center}
	\includegraphics[width=7.5cm ,clip]{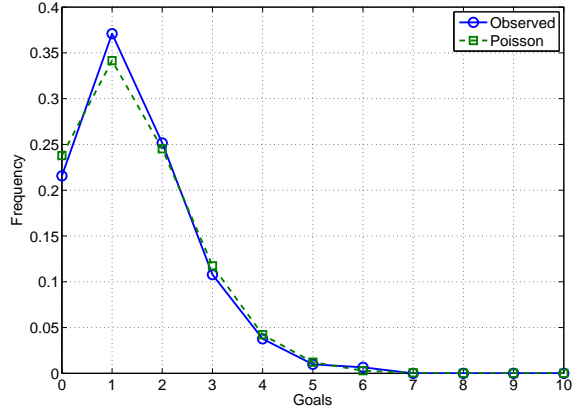}
	\caption{Distribution of goals per game}
	\label{fig:gpg_Poisson}
\end{center}
\end{figure}

It is well known that the number of goals in one football game follows a Poisson distribution\cite{raey}.
Figure \ref{fig:gpg_Poisson} shows the histogram of goals per game in the 2013 J1 League and the probability function of a  Poisson distribution with the actual average as the mean.

In this paper, the following models of the game are used.
\begin{itemize}
\item M1: $X_i$ is determined by  $Po(\lambda_{ALL})$ in all games.
In other words, 
\begin{equation}
p(x)=\displaystyle e^{-\lambda_{ALL}}\frac{\lambda_{ALL}^x}{x!}, x=0,1,\cdots.
\end{equation}

Further, suppose that the number of goals by the home and away teams are independent of each other:
\begin{equation}
P(X_i=x, X_j=y)=p(x)p(y).
\end{equation}

In other words, every team has the same offensive and defensive skills.

\item M2: $X_i$ is determined by  $Po(\lambda_{i,GF})$ for all opponents.
In other words, 
\begin{equation}
p_i(x)=\displaystyle e^{-\lambda_{i,GF}}\frac{\lambda_{i,GF}^x}{x!}, x=0,1,\cdots.
\end{equation}

Further, suppose that the number of goals by the home and away teams are independent of each other:
\begin{equation}
P(X_i=x, X_j=y)=p_i(x)p_j(y).
\end{equation}

In other words, every team has different offensive skills, but the same defensive skills.

\item M3: When $i$ plays against $j$, $X_i$ is determined by $\displaystyle Po\left(\frac{\lambda_{i,GF}+\lambda_{j,GA}}{2}\right)$.

\begin{equation}
p_{i,j}(x)=\displaystyle e^{-\mu_{i,j}}\frac{\mu_{i,j}^x}{x!}, x=0,1,\cdots.
\end{equation}
\begin{equation}
\mu_{i,j}=\frac{\lambda_{i,GF}+\lambda_{j,GA}}{2}.
\end{equation}

Further, suppose that the number of goals by the home and away teams are independent of each other:
\begin{equation}
P(X_i=x, X_j=y)=p_{i,j}(x)p_{j,i}(y).
\end{equation}

In this model, the offense and the defense skills are reflected.

\item M4: When $i$ plays against $j$ in $i$'s hometown, $X_i$ is determined by 

$\displaystyle Po\left(\frac{\lambda_{i,GF,H}+\lambda_{j,GA,A}}{2}\right)$.

\begin{equation}
p_{i,j}(x)=\displaystyle e^{-\mu_{i,j}}\frac{\mu_{i,j}^x}{x!}, x=0,1,\cdots.
\end{equation}
\begin{equation}
\mu_{i,j}=\frac{\lambda_{i,GF,H}+\lambda_{j,GA,A}}{2}.
\end{equation}

Further, suppose that the number of goals by the home and away teams are independent of each other:
\begin{equation}
P(X_i=x, X_j=y)=p_{i,j}(x)p_{j,i}(y).
\end{equation}

In this model, not only the offense and the defense skills but also the home and away status are reflected.

\item M5: 
Average of goals of $i$ against $j$ is estimated from the results of the past games.

\begin{itemize}
\item From the past results, team $i$ scored $g$ goals againt $j$ is added to detabase as $(\lambda_{i,GF}, \lambda_{j,GA}, g)$.
\item The axes of $\lambda_{i,GF}$ and $\lambda_{j,GA}$ are partitioned with interval $0.4$.
Average of the number of goals $g$ are calculated for each partition.
If there are less than 30 games in one partition, they are discarded as outlier.
\item Regression coefficients of the following regression model are estimated by least square method.
\begin{equation}
\mu_{i,j}\equiv \mu(\lambda_{i,GF}, \lambda_{j,GA})=a_1 \lambda_{i,GF}+ a_2 \lambda_{j,GA} +a_3
\end{equation}

\end{itemize}

Based on the regression model, when $i$ plays against $j$, $X_i$ is determined by $\displaystyle Po\left(\mu_{i,j}\right)$.

\begin{equation}
p_{i,j}(x)=\displaystyle e^{-\mu_{i,j}}\frac{\mu_{i,j}^x}{x!}, x=0,1,\cdots.
\end{equation}

Further, suppose that the number of goals by the home and away teams are independent of each other:
\begin{equation}
P(X_i=x, X_j=y)=p_{i,j}(x)p_{j,i}(y).
\end{equation}
\end{itemize}

The paremeters  $\lambda_{i,GF}$, $\lambda_{i,GA}$, and so on are taken from J1 League statistics.

\subsection{Simulation result}
For each model, $10^5$ seasons are simulated and the number of overlap cases is counted.
In M5, the regression alanysis is performed based on the results of 2010---2013 seasons.
$r^2=0.8721$.
Tables \ref{tab:numPlayoff} and \ref{tab:prOverlap} show the results.

\begin{table}[h]
\begin{center}
\caption{The number of teams reached the postseason}
\label{tab:numPlayoff}
\begin{tabular}{ccccccl} \\ \hline
Teams&M1&M2&M3&M4& M5&case {\#} \\ \hline
3&	0.4320  &0.5912 	&  0.5195	&	0.5676 &	0.6173 &  4, 6, 7, 8	\\ \hline
4&	0.4876 	&0.3740	&	0.4282	&  0.3916 &	0.3519 & 2, 3, 5\\ \hline
5&	0.0804	&0.0348	&	0.0522	&  0.0408 &	0.0309&1 \\ \hline \hline
mean&	3.6485&3.4466	&	3.5327	&	3.4372 &	3.4134\\ \hline
\end{tabular}
\end{center}
\end{table}

\begin{table}[h]
\begin{center}
\caption{Probability of each overlap cases}
\label{tab:prOverlap}
\begin{tabular}{ccccccc} \\ \hline
Teams&M1&M2&M3&M4&M5&case {\#}	\\ \hline
  5	&0.0804&	0.0348	&    0.0522	&	0.0408	& 0.0309&1	\\ \hline
	4	&0.0754&  0.0496	&    0.0606	&	0.0527	& 0.0459&2	\\ \hline
	4	&0.1205&  0.0865	&    0.1016	&	0.0898	& 0.0809&3	\\ \hline
	3	&0.0530&  0.0549	&    0.0539	&	0.0513	& 0.0550&4	\\ \hline
	4	&0.2917&  0.2379   &    0.2660	&	0.2491	& 0.2251&5	\\ \hline
	3	&0.1168&	0.1399	&    0.1328	&	0.1353	& 0.1448&6	\\ \hline
	3	&0.2063&	0.2766	&    0.2445	&	0.2639	& 0.2864&7	\\ \hline
	3	&0.0559&	0.1198	&    0.0873	&	0.1172	& 0.1311&8	\\ \hline
\end{tabular}
\end{center}
\end{table}    

\subsection{Discussion}
Table \ref{tab:numPlayoff} shows that a postseason with the maximum of five teams will be held only once about every 10 to 30 years.

\begin{figure}[h]
\begin{center}
	\includegraphics[width=9.0cm ,clip]{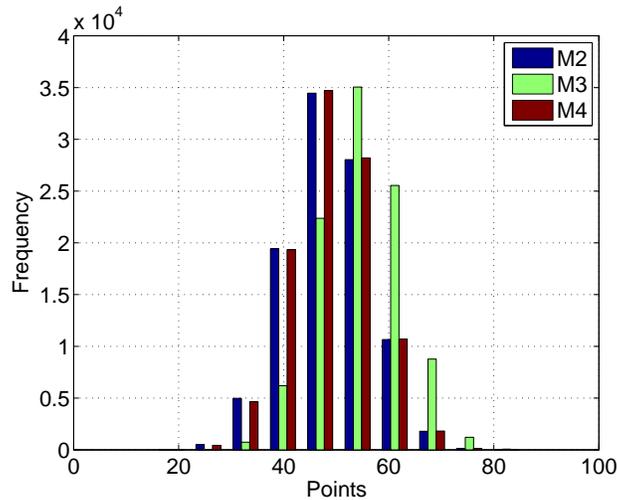}
	\caption{Distribution of points per season}
	\label{fig:dist_pps}
\end{center}
\end{figure}

For every model, the simulated total points are  
\begin{itemize}
\item less than the real values for the teams ranked high, and
\item more than the real values for the teams ranked low.
\end{itemize}
For instance, Figure \ref{fig:dist_pps} shows the simulated points obtained by  models M2 to M4 for the winning team in the 2013 season with 63 points.
The averages  are $48.705,   54.686,$ and $48.797$ by each model, respectively:
these are less then the actual score by 10 to 15 points.
This is a straightforward result expected from the calculation rule of average goals for (and goals against) in each model.
It can and should be modified in the future.
These models, however, lead the points distribution in a narrow range.
Therefore, a model close to the real dynamics may lead more overlaps.

M5 improves this point.
Table \ref{tab:resultM5} shows actual points (Pts), mean of simulated points (Mean)，error (Err)，standard deviation (Std), and normalized error (Err/Std).
The actual points of 16 out of 18 teams are within simulated $\mu\pm 1.2\sigma$.
This model fails prediction for the rest 2 teams by the regression analysis because 
there are few similar teams in terms of the goals for and against in the past results.
However, similar to M2 to M4,  M5 also leads the points distribution in a narrow range.
Therefore, a model close to the real dynamics may lead more overlaps. 

\begin{table}[htbp]
\begin{center}
\caption{Simulation result with model M5 (total points)}
\begin{tabular}{rrrrrr}
Standing&Pts&Mean&Err&Std&Err/Std\\ \hline
1 & 63 & 60.86  & $-2.14$ & 7.32  & $-0.292$\\ \hline
2 & 62 & 58.22  & $-3.78$ & 7.37  & $-0.512$\\ \hline
3 & 60 & 54.64  & $-5.36$ & 7.67  & $-0.698$\\ \hline
4 & 59 & 59.92  & 0.92 & 7.39  & 0.124 \\ \hline
5 & 59 & 51.28  & $-7.72$ & 7.63  & $-1.011$\\ \hline
6 & 58 & 52.24  & $-5.76$ & 7.71  & $-0.747$\\ \hline
7 & 55 & 50.41  & $-4.59$ & 7.53  & $-0.609$\\ \hline
8 & 54 & 54.86  & 0.86 & 7.63  & 0.112 \\ \hline
9 & 50 & 41.67  & $-8.33$ & 7.47  & $-1.115$\\ \hline
10 & 48 & 45.16  & $-2.84$ & 7.63  & $-0.372$\\ \hline
11 & 47 & 46.12  & $-0.88$ & 7.52  & $-0.117$\\ \hline
12 & 46 & 41.85  & $-4.15$ & 7.55  & $-0.549$\\ \hline
13 & 45 & 48.65  & 3.65 & 7.41  & 0.492 \\ \hline
14 & 45 & 44.90  & $-0.10$ & 7.47  & $-0.013$\\ \hline
15 & 37 & 39.43  & 2.43 & 7.10  & 0.342 \\ \hline
16 & 25 & 30.72  & 5.72 & 6.90  & 0.828 \\ \hline
17 & 23 & 37.37  & 14.37 & 7.22  & 1.990 \\ \hline
18 & 14 & 26.62  & 12.62 & 6.55  & 1.926 \\ \hline
\end{tabular}
\label{tab:resultM5}
\end{center}
\end{table}

The official definition of Super Stage states the following:
\begin{quote}
	The Super Stage is a knockout tournament of four teams.
	The following four teams reach the Super Stage.
		\begin{itemize}
		\item Winning teams of the 1st and 2nd stages.
		\item The teams ranked 2nd and 3rd in terms of  total points over the two stages.
		\end{itemize}
\end{quote}
This definition does not properly reflect the facts of high probability of overlap.
Practically, this postseason is basically by Y1, Y2 and Y3.
If a stage-winning team is (unfortunately) ranked 4th or lower can reach the postseason as a repechage team.
This paper proposes the following (proper) definition of the postseason.
Note that this definition proposes only proper explanation but not proposes proper design of postseason system. 

{\bf Definition of the postseason of J1 League from 2015 season}

\begin{itemize}
\item Regular season
	\begin{itemize}
	\item Ragular season consists of two stages by 18 clubs.
	Each stage is named as ``1st stage" and ``2nd stage."
	\item Each stage is single round-robin competition. Home and away games are placed on either stage. 
	As a result, each team has 17 matches in one stage with 8 or 9 home games.
	\item Standings for each stage are based on the sum in points of each game.
	The point system is the usual 3-1-0 (Win-Draw-Lose) system.
	\end{itemize}

\item Postseason
	\begin{itemize}
		\item The final of the postseason tournament is called ``Championship", and the rest is ``Super Stage".
		
	\item The teams ranked 1st, 2nd, and 3rd in terms of  total points over the two stages reach the postseason tournament. They are denoted as Y1, Y2, and Y3.
	Y1 is seeded to the Championship.
	Y2 and Y3 play in the Super Stage.
	The winner of the Super Stage reaches the Championship
	(Fig. \ref{fig:JPlayoff3Teams}).
	\begin{figure}[h]
\begin{center}
	\includegraphics[width=9.0cm ,clip]{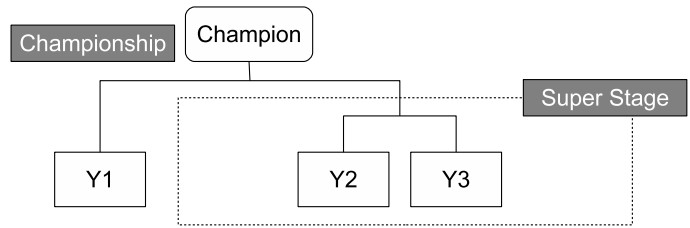}
	\caption{Basic format of postseason}
	\label{fig:JPlayoff3Teams}
\end{center}
\end{figure}
	
	\item If a stage-winning team is ranked 4th or lower (denoted as Y4-) can reach the postseason.
	The stage-winning teams are denoted by W1 and W2 based on {\bf their total points over both stages}, not based on the stage number.
		\begin{itemize}
		\item If W1 is in Y1 to Y3 and W2 is in Y4-, W2 reaches the first round.
		
			\begin{itemize}
			\item If W1 is Y2 or Y3, W1 is seeded and clears the first round (Fig. \ref{fig:JPlayoff4Teams}(a)).
			\item If W1 is Y1, W2 is seeded and clears the first round. Y2 and Y3 play in the first round (Fig. \ref{fig:JPlayoff4Teams}(b)).
			\end{itemize}
			
			\begin{figure}[h]
\begin{center}
	\includegraphics[width=9.0cm ,clip]{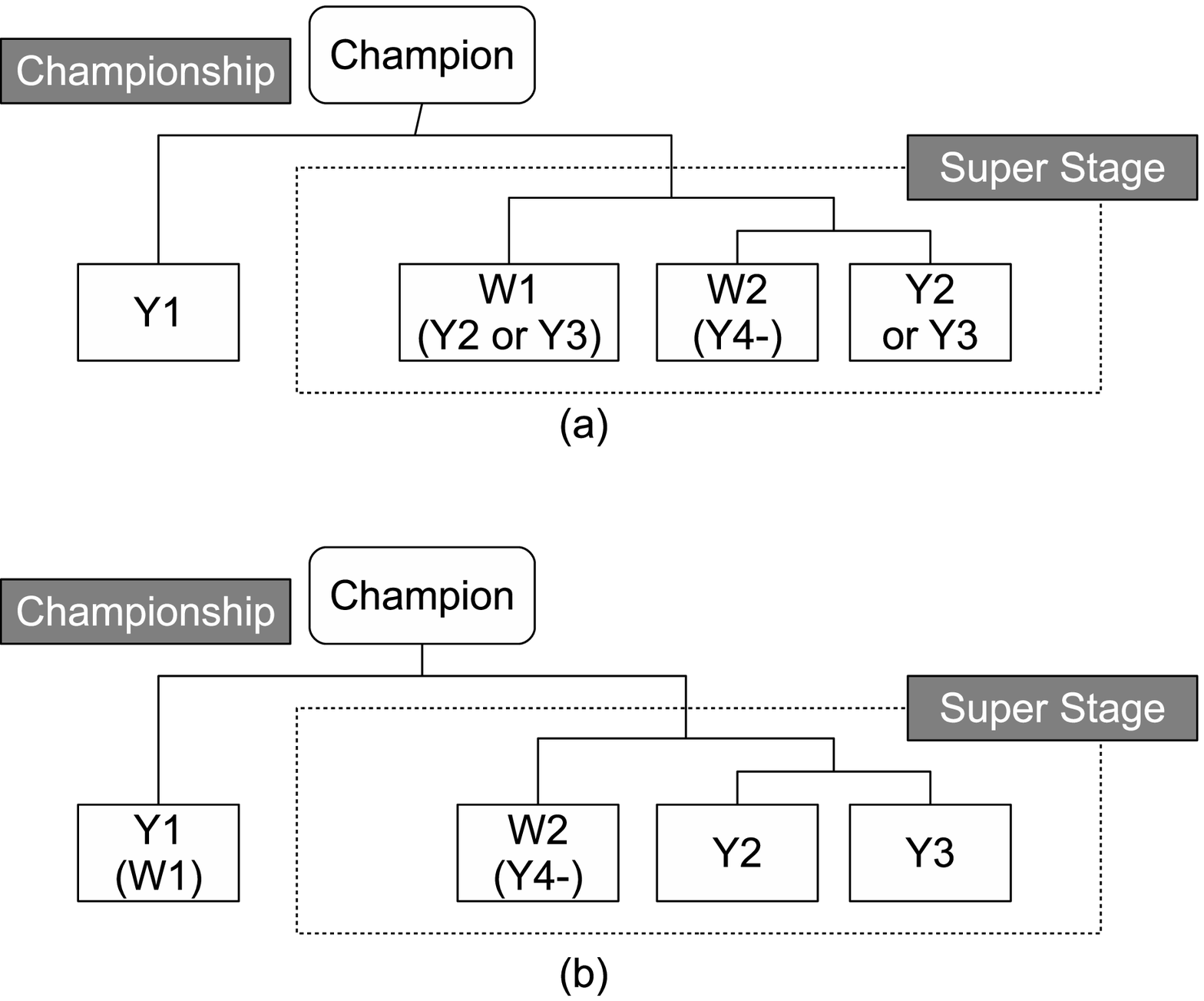}
	\caption{One stage winner reaches postseason as repechage}
	\label{fig:JPlayoff4Teams}
\end{center}
\end{figure}

		\item If both W1 and W2 is in Y4-, they reach the postseason.
		The first round matches are Y2 v. W2, and Y3v. W1
		(Fig. \ref{fig:JPlayoff5Teams}).

			\begin{figure}[h]
\begin{center}
	\includegraphics[width=9.0cm ,clip]{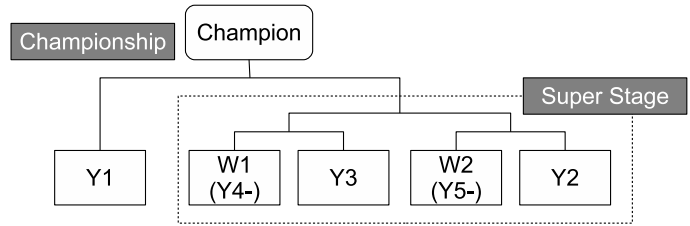}
	\caption{Two stage winners reach postseason as repechage}
	\label{fig:JPlayoff5Teams}
\end{center}
\end{figure}

		\end{itemize}
	\end{itemize}
\end{itemize}

This definition is much simpler than the official one.
And this definition clarifies that the new season system from 2015 leads a one-stage regular season with the postseason tournament, whereas the new system is officially defined as a two-stage system, due to inappropriate design of the postseason.

\section{Conclusion}

From these reuslts, it can be concluded that the new J1 postseason system from the 2015 season has inherent overlaps.
In addition, it is clarified that the new sytem is basically a one-stage system with the postseason tournament, 
and it  lowers the prestige of winning the 1st and 2nd stages because these teams would appear postseason as repechage teams.
Given the above, it would be interesting to see whether the new  league competition setup succeeds in attracting increasing attention thereby raising sales and profits.

In the future, we intend to develop a more accurate game model and calculate the overlap probability.


\end{document}